\documentclass[12pt,letterpaper,copyedit]{article}
\usepackage{osajnl2}

\usepackage{amsmath}

\begin{document}


\title{Optical Realization of Coherent Vibrational Dynamics in Molecules}


\author{Stefano Longhi}
\address{Dipartimento di Fisica and Istituto di Fotonica e Nanotecnologie del CNR,
Politecnico di Milano, Piazza L. da Vinci 32, I-20133 Milano, Italy}
\begin{abstract}
Optical analogues of coherent vibrational phenomena in molecules,
such as light-induced molecular stabilization and wave packet dynamics at a potential crossing, are
proposed for light beams in coupled slab waveguides.
\end{abstract}

\ocis{130.2790, 230.3120, 230.7370, 000.1600}



\noindent

Light propagation in coupled optical waveguides provides an
excellent laboratory tool to visualize with classical waves the
dynamical aspects embodied in a wide variety of coherent phenomena
encountered in atomic, molecular or condensed-matter physics
\cite{Longhi09LPR}. Among others, we mention adiabatic stabilization
of atoms in ultrastrong laser fields \cite{Longhi05}, Bloch
oscillations, Zener tunneling and dynamic localization in
crystalline potentials \cite{Peschel98,Trompeter06,Longhi06},
coherent population transfer in laser-driven atoms
\cite{Longhi07,Lahini08}, and quantum Zeno dynamics
\cite{Biagioni08}. As compared to quantum systems, optics offers
unique possibilities such as a direct visualization in space of
typical ultrafast phenomena in time and the ability to explore
coherent dynamical regimes of difficult access in quantum systems or
obscured by detrimental effects such as dephasing and
particle-particle interaction. Previous optical analogies
\cite{Longhi09LPR} deal with coupled channel waveguides in an
assigned geometry, in which spatial light transport turns out to be
analogous to the coherent evolution of a single-particle electronic
wave function. Unfortunately, these
 photonic structures could not account for
vibrational effects typical of e.g. molecules or crystals. Nuclear
degrees of freedom in molecules are known to yield a wealth of
interesting phenomena, especially when driven by ultrastrong laser
fields. Coherent effects in molecules, like laser-assisted
preparation and control of vibrational wavepackets and molecular
bond engineering, provide theoreticians and experimentalists  with
examples of time-dependent quantum mechanics at work, with important
applications to femtochemistry
\cite{Garraway95,Yeazell00,Posthumus04}. Though current pump-probe
techniques enable the observation of time-resolved wave packet
dynamics of nuclear motion, optical analogues of coherent
vibrational phenomena may offer the possibility to assess wave
packet dynamical phenomena of difficult access in the quantum realm,
such as wave packet dynamics and momentum filtering at surface
potential crossings
\cite{Suominent93}.\\
In this Letter optical analogues of laser-assisted coherent
vibrational phenomena in molecules, such as light-induced molecular
stabilization and wave packet dynamics at surface potential
crossings, are theoretically proposed for light beams propagating in
coupled slab waveguides. The photonic structure, schematically shown
in Fig.1(a), consists of two slab waveguides S$_1$ and S$_2$, placed
at a distance $d$, which confine the field in the transverse $y$
direction. In the weak guidance approximation, light propagation at
wavelength $\lambda$ is described by the Schr\"{o}dinger-like wave
equation for the electric field envelope $\psi(x,y,z)$
\cite{Longhi09LPR,Sukhorukov03}
\begin{equation}
i \hbar \partial_z \psi=-[\hbar ^2/(2 n_s)]
\nabla^2_t \psi+[n_s-n(x,y)] \psi
\end{equation}
 where $\nabla^2_t=\partial^2_x+\partial^2_y$ is the transverse Laplacian,
 $\hbar= \lambda / (2 \pi)=1/k$ is the reduced wavelength, $n_s$ is
the bulk refractive index, $n=n_s+\Delta n_1(x,y-d/2)+\Delta
n_2(x,y+d/2)$ is the refractive index profile of the structure, and
$\Delta n_{1,2}(x,y)$ are the refractive index increase in the two
slab guiding regions. The index profiles $\Delta n_{1,2}$ are
generally allowed to slowly vary along the horizontal $x$ direction,
such as in case of graded-index slabs or slabs with slowly-varying
thickness  \cite{Sukhorukov03}. We indicate by $q_1(y;x)$ and
$q_2(y;x)$ the fundamental mode profiles of the two slabs at a given
transverse coordinate $x$ and by $W_{1,2}(x)$ the respective
effective indices, i.e. $-(\hbar^2/2n_s) \partial^2_y q_{1,2}+\Delta
n_{1,2}(x,y)q_{1,2}= W_{1,2}q_{1,2}$. Assuming that $W_{1,2}(x)$
differ from a reference value $W_0$ by a small quantity (of order
$\sim \epsilon^2$) and  that the field $\psi$ and the effective
indices $W_{1,2}$ vary slowly with respect to $x$, a multiple-scale
asymptotic analysis shows that the solution to Eq.(1) can be written
as $\psi(x,y,z)=[\psi_1(x,z) q_1(y-d/2;x)+\psi_2(x,z) q_2(y+d/2;x)]
\exp(-i W_0 z/ \hbar)+O(\epsilon^2)$, where the slowly-varying
amplitudes $\psi_{1,2}$ satisfy the coupled Schr\"{o}dinger-like
equations
\begin{equation}
i \hbar \partial_z \psi_{1,2} = -[\hbar^2/(2n_s)] \partial^2_x
\psi_{1,2}+ \mathcal{U}_{1,2}(x) \psi_{1,2}+ \kappa(x) \psi_{2,1}.
\end{equation}
In Eqs.(2), we have set $\mathcal{U}_{1,2} = W_{1,2}(x)-W_0+\int dy
\Delta n_{2,1} q_{1,2}^2+(\hbar^2/2n_s) \int dy (\partial_x
q_{1,2})^2
 \simeq  W_{1,2}(x)-W_0$, $\kappa  =  \int dy \Delta n_1 q_1 q_2=\int dy \Delta n_2 q_1 q_2$
and the normalization conditions $\int dy q_{1,2}^2=1$ have been
assumed. In their present form, Eqs.(2) are analogous to the quantum
mechanical equations describing the vibrational dynamics of a
molecule onto two energy surfaces $\mathcal{U}_{1,2}$ coupled by an
external laser field (interaction strength $\kappa$) in the
Born-Oppenheimer and rotating-wave approximations (see, for
instance, \cite{Garraway95,Suominent93}). Note that the temporal
variable in the quantum problem is replaced here by the spatial
propagation distance $z$, the reduced nuclear mass by $n_s$, the
Planck constant by the wavelength $\lambda$, and the inter-nuclear
distance by the coordinate $x$. This analogy enables to investigate
in optics the classical analogues of a wide variety of wave packet
phenomena encountered in molecular systems. We will limit here to
discuss two of such analogies. The first one, shown in Fig.1(b), is
the phenomenon of laser-induced stabilization of a dissociating
molecule, which is usually explained by the formation of
light-induced adiabatic surface potentials $\mathcal{U}_{\pm}(x)$,
defined by the equation
$\mathcal{U}_{\pm}^2-(\mathcal{U}_1+\mathcal{U}_2) \mathcal{U}_{\pm}
-\kappa^2+\mathcal{U}_1 \mathcal{U}_2=0$ (see, for instance,
\cite{Garraway95,Wunderlich96,Frasinski99}). Let us consider a
diatomic molecule (like H$_2^+$) characterized by a bonding
($\mathcal{U}_1$) and an antibonding ($\mathcal{U}_2$) potential
curve. A dissociating vibrational wave packet is initially created
from the ground vibrational state at the internuclear distance $x_1$
by a strong and ultrashort laser pulse at frequency $\omega_1$
[Fig.1(b)]. If a second continuous-wave laser field at frequency
$\omega_2$ is applied to set in resonance the two potential curves
at a distance $x_2>x_1$, a potential well is formed on the adiabatic
surface  $\mathcal{U}_{+}(x)$ at the avoided crossing [see
Fig.1(b)], and molecular dissociation in thus inhibited in the
adiabatic limit. The nuclear distance $x_2$ at which the new bond is
formed can be tuned by varying the frequency $\omega_2$ of second
laser. The optical realization of such a process is shown in
Fig.2(a). Here the two slab waveguides have a slowly-varying
thickness, linearly decreasing with $x$ for slab S$_1$ and linearly
increasing for slab S$_2$. The slabs are truncated at $x=x_0$,
providing a steep potential barrier for light. An equal slab
thickness is reached at $x=x_2>x_0$. Correspondingly, the potential
curves $\mathcal{U}_1$ and $\mathcal{U}_2$ entering in Eqs.(2) mimic
 the bonding and anti-bonding potential curves of Fig.1(b), with a level crossing at $x=x_2$. Moreover,
 the coupling rate $\kappa$ turns out to be nearly independent
 of $x$. In our waveguide system,
 excitation of a dissociating wave packet at $x=x_1$ on the antibonding surface is simply achieved by launching an elliptic
 Gaussian-shaped beam into the slab S$_2$ at normal incidence [inset in Fig.2(a)].
 Inhibition of molecular dissociation, arising from the formation of an adiabatic trapping well
 as the coupling strength $\kappa$ increases, is shown in Figs.2(b)-(d). The integrated intensity
  light distributions $\int_{{\rm S}_1}|\psi(x,y,z)|^2dy \propto |\psi_1(x,z)|^2$
and $\int_{{\rm S}_2}|\psi(x,y,z)|^2dy \propto |\psi_2(x,z)|^2 $ trapped in the two slabs, shown in Fig.2(b)-(d),
are numerically computed by a standard
beam propagation analysis of Eq.(1). Note that, for a large waveguide separation [Fig.2(b)], the wave packet
on the antibonding curve is accelerated toward increasing values of $x$ and the optical analogue of molecular dissociation is attained.
As the waveguide separation decreases (and thus the coupling $\kappa$ increases),
wave packet trapping is clearly observed [Fig.2(d)]. In this case, at the crossing point $x=x_2$ the wave packet
periodically tunnels between the two waveguides, and is accelerated in opposite directions owing to the opposite
slopes of the curves $\mathcal{U}_1$ and $\mathcal{U}_2$. This trapping mechanism is imperfect due to Landau-Zener tunneling at each
level crossing, and the lifetime of the trapped oscillatory motion increases as $\kappa$
increases [compare Figs.2(c) and (d)].\\
The second analogy is the acceleration of a molecular wave packet
after a potential crossing due to the dependence of the Landau-Zener
transition probability on the wave packet velocity
\cite{Suominent93}. Let us consider a wave packet that propagates,
at a constant speed $v_0$, along a flat potential curve
$\mathcal{U}_1(x)=0$ and undergoing a Landau-Zener crossing with a
 second potential curve $\mathcal{U}_2(x)=F(x-x_2)$ at $x=x_2$ [see Fig.1(c)].
 In the semiclassical approximation, the wave packet behaves like a particle with a definite momentum $p_0=n_s v_0$, which is assumed
 to weakly spread during the crossing and to keep its initial velocity $v_0$. After the crossing,
the probability $P_1$ of the wave packet to remain on the original potential surface $\mathcal{U}_1$ is
given by the usual Landau-Zener formula \cite{Garraway95,Suominent93} $P_1=\exp(-\pi \Lambda)$, where
$\Lambda=2 \kappa^2/(|F| \hbar v_0)$. Note that $P_1$ is larger for a faster wave packet.
If the wave packet momentum distribution is broad around its mean $p_0$, a more complex scenario is found.
 In particular, owing to the dependence of $P_1$ on $v_0$, a filtering effect
in the momentum space occurs, with the faster components of the wave
packet preferentially remaining on the original potential curve. The
final result is a slight increase of the mean wave packet velocity
\cite{Suominent93}. The analogue of this phenomenon for light beams
is shown in Fig.3. Here two slab waveguides, the first one with
constant and the second one with a linearly-increasing thickness
[Fig.3(a)], mimic the energy crossing scenario of Fig.1(c) with a
force $F \simeq 34 \; \mu {\rm m}/ {\rm cm}^{2}$ and a coupling
$\kappa / \hbar \simeq 1.04 \; {\rm cm}^{-1}$. A tilted
Gaussian-shaped wave packet
$\psi(x,y,0)=\exp\{-[(y-d/2)/w_y]^2-[(x-x_1)/w_x]^2-i v_0 n_s k x
\}$, with $w_x= 8 \; \mu$m, $w_y= 2.4 \; \mu$m, $d= 10.7 \; \mu$m,
and $x_1= -800 \; \mu$m, excites waveguide S$_1$ with an initial
transverse velocity (refraction angle) $v_0=278 \; \mu$m/cm,
corresponding to a Landau-Zener probability $P_1 \sim 0.48$. Figure
3(b) depicts the path followed by the center of mass $\langle x
\rangle$  of $\psi_1$ (solid line); for comparison, the straight
path followed by $\psi_1$ in absence of slab S$_2$ is also depicted
(dotted curve). The two insets in Fig.3(b) show the the evolution of
the fractional beam power $P_1(z)$ of wave packet $\psi_1$ and its
instantaneous velocity $v(z)=d\langle x \rangle /dz$. Note that, as
the wave packet reaches the crossing region and light transfer to
S$_2$ starts, the velocity $v(z)$ first decreases from the initial
value $v_0$, then increases and reaches an asymptotic value slightly
larger than $v_0$, indicating an acceleration of the wave packet.
Such a behavior can be explained \cite{Suominent93} after observing
that the faster wave packet components reach first the crossing
region and partially tunnel into slab S$_2$, so that a deceleration
of $\psi_1$ is initially observed. When the slower wave packet
components reach the crossing regions, they are partially
transferred into slab S$_2$ and an acceleration of $\psi_1$ is thus
observed. Since the tunneling probability is smaller for the faster
wave packet components, the final mean velocity of $\psi_1$ is
larger than the launching one $v_0$. This effect, however, is small
because the change of beam path remains internal to the diffraction
cone of the beam [see the shaded  area in Fig.3(b)].
Similar behavior is obtained by reversing the slope of the potential $\mathcal{U}_2$. \\
Optical analogs of coherent phenomena in matters have so far
visualized the behavior of the electronic wave function solely
\cite{Longhi09LPR}.
 Here a photonic structure capable of mimicking coherent vibrational effects in molecules has been proposed.
The present results may stimulate theoretical and experimental
studies on optical analogs of coherent vibrational phenomena in molecules or crystals.

 Author E-mail address: longhi@fisi.polimi.it

\newpage
{\bf List of Figure captions}\\
\\
{\bf Fig.1} (Color online) (a) Schematic of a two-slab optical
waveguide structure. (b) Nuclear potential energy diagram of a
diatomic molecule interacting with two laser fields. (c) Schematic
of wave packet splitting at a linear potential energy crossing.\\
\\
{\bf Fig.2} (Color online) (a) Refractive index of the slab
waveguide structure that mimics the molecular energy diagram of
Fig.1(b). The inset shows the elliptical Gaussian beam that excites
guide S$_2$ at $x=x_1$. (b)-(d): Evolution of light distributions
$|\psi_1(x,z)|^2$ (right panels) and $|\psi_2(x,z)|^2$ (left panels)
for decreasing values of waveguide separation $d$: (b) $d=12 \;
\mu$m, (c) $d=9 \; \mu$m, (d) $d=8 \; \mu$m. Other parameters are:
$\lambda=633$ nm, $n_s=1.45$.\\
\\
{\bf Fig.3} (Color online) Wave packet dynamics at a potential
crossing. (a) Refractive index of the slab waveguides that mimics
the potential crossing of Fig.1(c). (b) Path followed by the center
of mass of wave packet $\psi_1$ (solid curve). The shaded area shows
beam spreading. The insets show the evolution of fractional beam
power $P_1$ (bottom) and speed $v$ (top) of wave packet $\psi_1$.

\newpage

\begin{figure}[htb]
\centerline{\includegraphics[width=8.2cm]{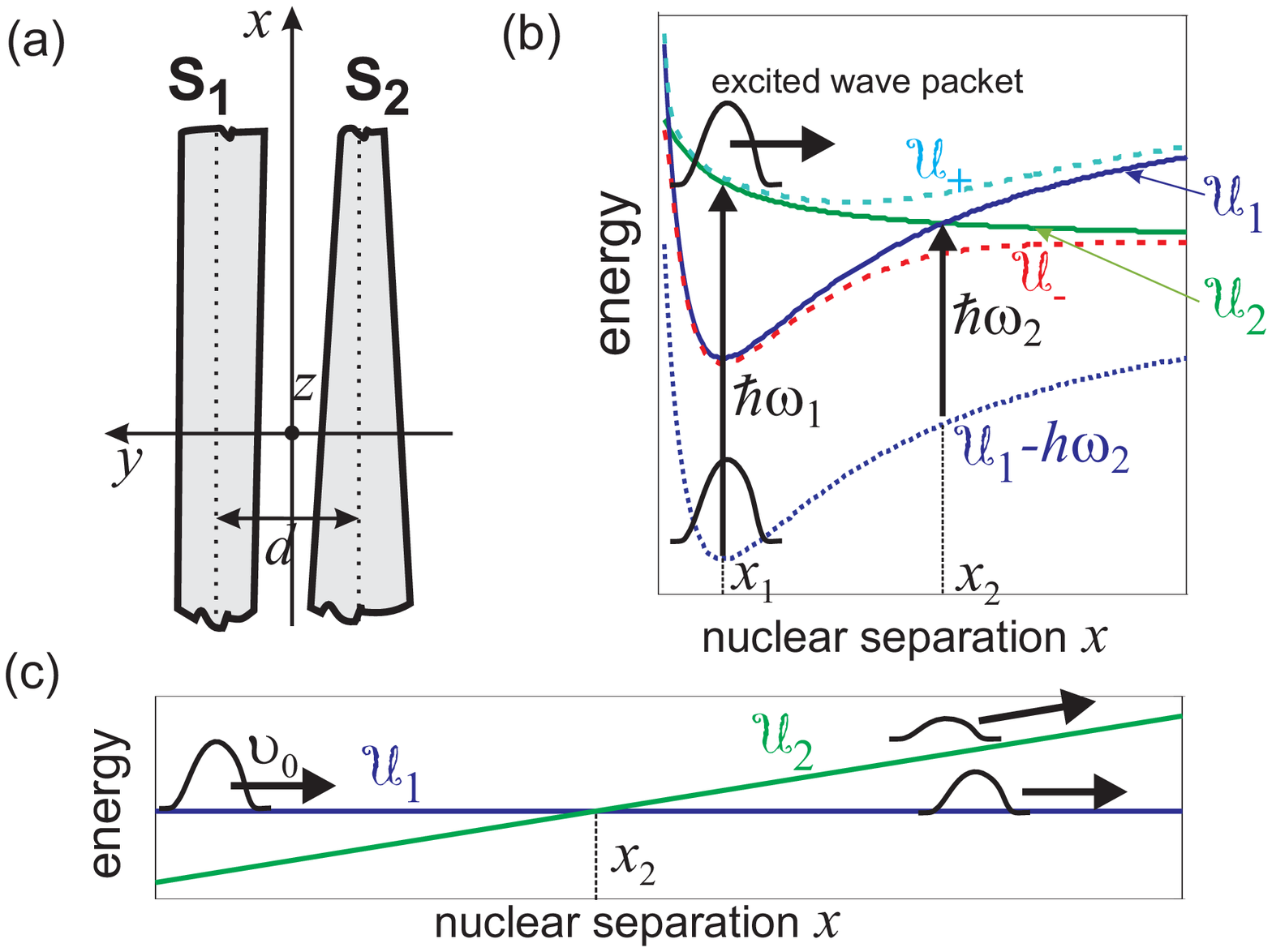}} \caption{}
\end{figure}

\newpage

\begin{figure}[htb]
\centerline{\includegraphics[width=8.2cm]{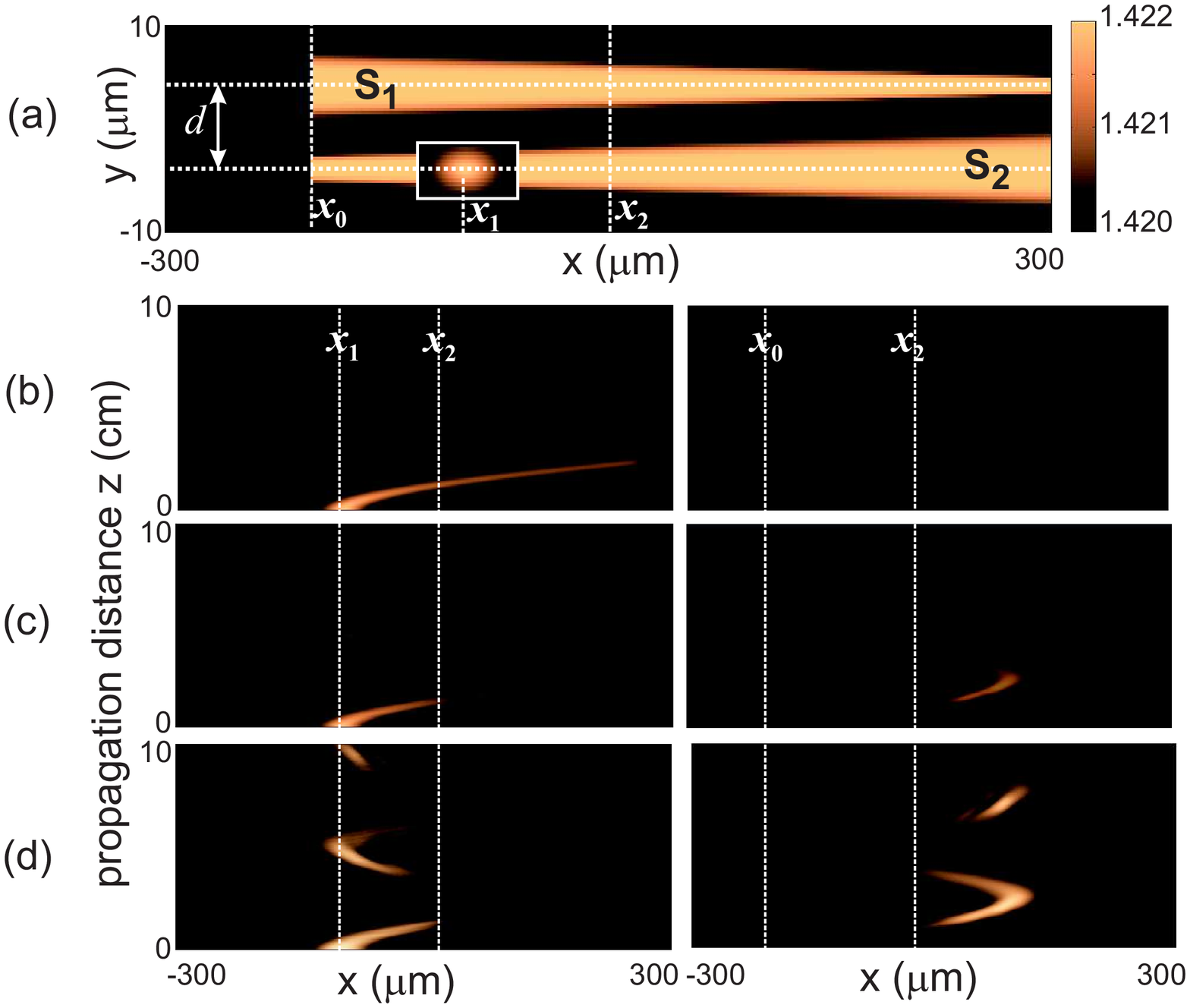}} \caption{}
\end{figure}

\newpage

\begin{figure}[htb]
\centerline{\includegraphics[width=8.2cm]{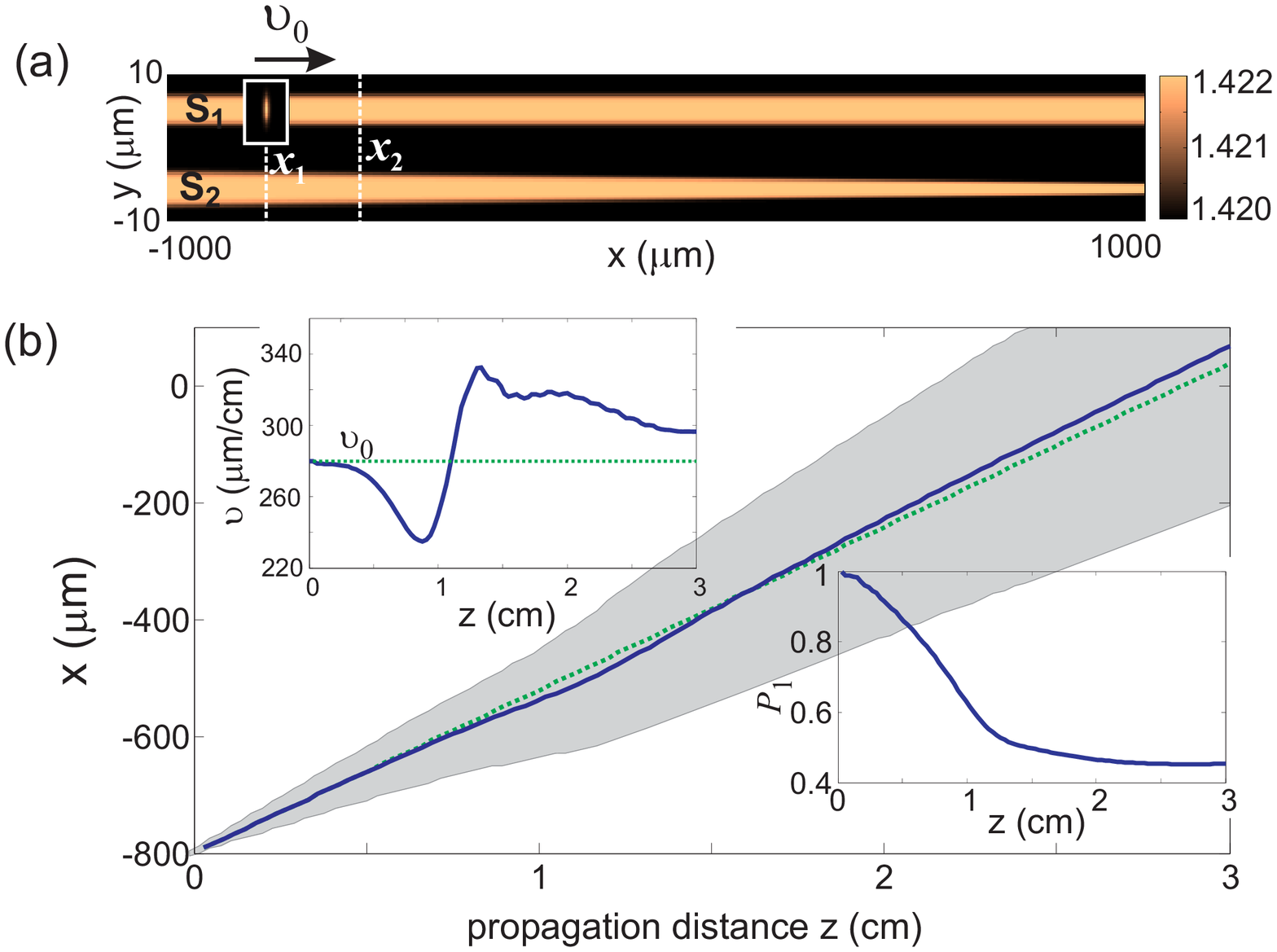}} \caption{}
\end{figure}

\end{document}